\documentclass[journal=jctcce,manuscript=article]{achemso}

\usepackage[version=3]{mhchem} 
\usepackage{xcolor}
\usepackage{soul}
\usepackage{siunitx}
\usepackage{setspace}
\usepackage{wrapfig}

\usepackage{hyperref} 
\author{Ofir Blumer}
\affiliation[TAU]
{School of Chemistry, Tel Aviv University, Tel Aviv 6997801, Israel.}
\author{Barak Hirshberg}
\email{hirshb@tauex.tau.ac.il}
\affiliation[TAU]
{School of Chemistry, Tel Aviv University, Tel Aviv 6997801, Israel.}
\alsoaffiliation[Sackler]
{The Center for Computational Molecular and Materials
Science, Tel Aviv University, Tel Aviv 6997801, Israel.}
\alsoaffiliation[BioSoft]
{The Center for Physics and Chemistry of Living Systems, Tel Aviv University, Tel Aviv 6997801, Israel.}

\title[]{Have you tried turning it off and on again? Stochastic resetting for enhanced sampling}


\begin{document}

\begin{abstract}

Molecular dynamics simulations are widely used across chemistry, physics, and biology, providing quantitative insight into complex processes with atomic detail. However, their limited timescale of a few microseconds is a significant obstacle in describing phenomena such as conformational transitions of biomolecules and polymorphism in molecular crystals.
Recently, stochastic resetting, i.e., randomly stopping and restarting the simulations, emerged as a powerful enhanced sampling approach, which is collective variable-free, highly parallelized, and easily implemented in existing molecular dynamics codes. Resetting expedites sampling rare events while enabling the inference of kinetic observables of the underlying process. It can be employed as a standalone tool or in combination with other enhanced sampling methods, such as Metadynamics, with each technique compensating for the drawbacks of the other. Here, we comprehensively describe resetting and its theoretical background, review recent developments in stochastic resetting for enhanced sampling, and provide instructive guidelines for practitioners.

\end{abstract}

\singlespacing

\begin{figure}
    \centering
   \includegraphics[width=0.5\linewidth]{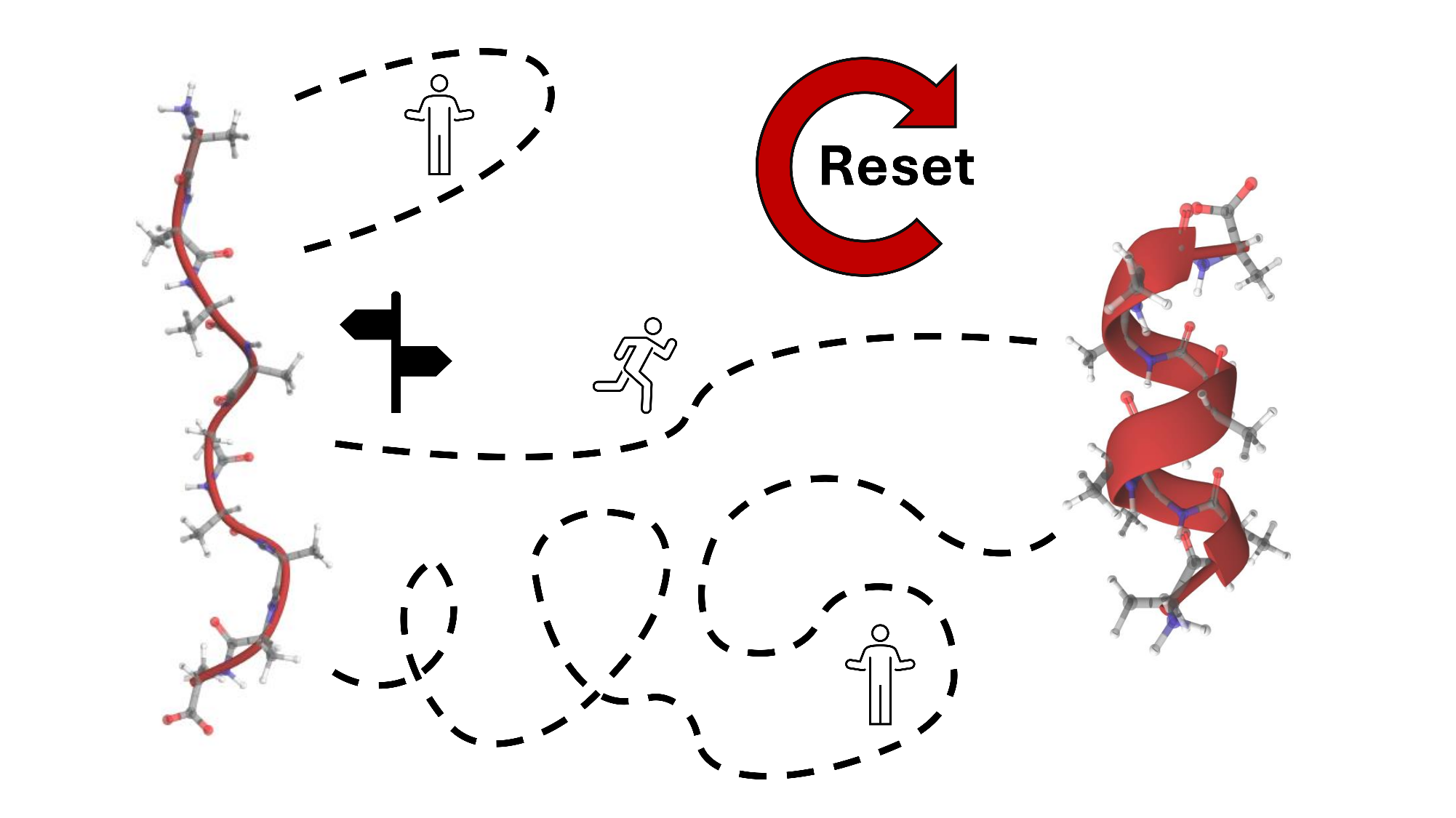}
    \label{fig:TOV}
\end{figure}
\noindent Molecular dynamics describes complex processes in microscopic detail but is limited in timescale. Occasionally restarting the simulations can significantly enhance the sampling of rare events and overcome this timescale problem.

\newpage
\section{Introduction}

Molecular dynamics (MD) simulations follow the motion of chemical, physical, and biological systems with atomic resolution. They provide valuable insights into the mechanism of complex processes and allow quantitative estimates of thermodynamic and kinetic properties. However, their inherent microscopic resolution dictates simulation time steps on the order of a femtosecond, limiting their overall timescale to a few microseconds. Processes that occur on longer timescales, such as protein conformational transitions and crystal nucleation, growth, and polymorphism, cannot be studied directly in standard MD simulations, and require specialized enhanced sampling algorithms~\cite{yang_thermodynamics_2015,Henin_Lelievre_Shirts_Valsson_Delemotte_2022,Kleiman_addaptive_2023,ray_kinetics_2023,Blumer2022}.

Different enhanced sampling methods were developed over the years, such as umbrella sampling~\cite{torrie_nonphysical_1977,kastner_umbrella_2011}, replica-exchange~\cite{sugita_replica-exchange_1999}, free-energy dynamics~\cite{rosso_adiabatic_2002,rosso_use_2002}, Gaussian accelerated MD~\cite{doi:10.1021/acs.jctc.5b00436}, on-the-fly probability enhanced sampling (OPES)~\cite{invernizzi_rethinking_2020,invernizzi_opes_2021}, milestoning~\cite{faradjian_computing_2004,elber_milestoning_2020}, transition path sampling~\cite{dellago1998transition,bolhuis2002transition}, forward flux sampling~\cite{Allen_2009}, transition interface sampling~\cite{VanErp20037762}, weighted ensemble~\cite{HUBER199697,doi:10.1146/annurev-biophys-070816-033834}, and
Metadynamics (MetaD)~\cite{barducci_metadynamics_2011,valsson_enhancing_2016,sutto_new_2012,bussi_using_2020}.
Many of them, and in particular MetaD and its variants~\cite{barducci_well-tempered_2008,valsson2014variational,ray_rare_2022}, rely on identifying efficient collective variables (CVs). These are functions of phase space variables that can distinguish between metastable states and describe the dynamics between them~\cite{demuynck_protocol_2018}. Good CVs provide high accelerations and accurate estimations of thermodynamic and kinetic properties but are very challenging to obtain~\cite{sidky_machine_2020}. Despite much progress in recent years~\cite{peters2006obtaining,mendels_collective_2018,bonati_deep_2021,sidky_machine_2020,chen_collective_2021,liu_graphvampnets_2023,doi:10.1021/acs.jctc.3c00051}, often only suboptimal CVs are available, leading to limited accelerations and slow convergence~\cite{invernizzi_making_2019,dietschreit_entropy_2023,Blumer2024}.

We recently developed a CV-free enhanced sampling method based on stochastic resetting (SR):~\cite{Blumer2022,Blumer2024,non_exponential_kinetics,church2024acceleratingmoleculardynamicsinformed} stopping the simulations at random times and restarting them, resampling the initial conditions. It can expedite the sampling of rare events
by an order of magnitude when used as a standalone tool~\cite{Blumer2022,non_exponential_kinetics}, and accelerate MetaD simulations by more than two orders of magnitude~\cite{Blumer2024}. It enables the inference of unbiased kinetics from accelerated simulations~\cite{Blumer2022, Blumer2024, church2024acceleratingmoleculardynamicsinformed}, even for non-exponential processes~\cite{non_exponential_kinetics}, and very recently inspired an improved kinetics inference procedure for MetaD simulations without resetting~\cite{blumer2024shorttime}.

One of the most appealing features of SR as an enhanced sampling method is 
its incredible simplicity: standard resetting requires nothing more than stopping a simulation and restarting it while continuing monitoring the overall simulation time. Resetting can be implemented straightforwardly in existing codes, and combined with any other enhanced sampling method. As trajectories with SR are composed of statistically independent segments, resetting is also highly parallelized. 
Resetting is useful because it is based on a rigorous theory, which allows predicting all the observables of the process under resetting from trajectories without resetting. For instance, it provides a universal sufficient criterion (which we describe in detail later) to test whether additional sampling of rare events can be achieved through simulations with resetting, and the resulting speedup.  

Here, we thoroughly review stochastic resetting for enhanced sampling. We first define what SR is in Section \ref{sec:whatis}. We then summarize key results in the field of SR in Section \ref{sec:theory}, which give sufficient conditions to when a stochastic process will be accelerated by resetting. Next, we discuss the applications of resetting as a standalone method and in combination with MetaD to accelerate MD simulations. We also present the concept of informed resetting, where we exploit information on reaction progress for greater speedups~\cite{church2024acceleratingmoleculardynamicsinformed}. Section \ref{sec:kinetics} reviews procedures to infer the unbiased kinetics of the system from accelerated simulations with resetting, with and without MetaD. Section \ref{sec:lookout} overviews anticipated future development of the resetting method. We conclude the review with a practical discussion of when and how should practitioners use resetting for enhanced sampling.

\section{What is stochastic resetting?}
\label{sec:whatis}

Before reviewing the theory of resetting in detail, and outlining sufficient and universal conditions for accelerating a stochastic process, we qualitatively discuss the typical behavior of stochastic processes that are accelerated under resetting.

Resetting is the procedure of occasionally stopping and restarting a random process, subject to the resampling of independent and identically distributed initial conditions. It has drawn much attention for over a decade,~\cite{Evans_majumdar_JPhysA_review,kundu2024preface} since the pioneering work of Evans and Majumdar~\cite{Evans_majumdar_PRL}. They studied a particle diffusing in one dimension, which is returned to its initial position (reset) at random times, sampled from an exponential distribution with a fixed rate $r$. Evans and Madjumadar showed this system is fundamentally different than the corresponding process without resetting: While the position distribution without resetting is a Gaussian with time-dependent width, under resetting the system reaches a steady state with a Laplace position distribution. Furthermore, the mean time for the particle to reach an absorbing boundary at a distance $L$ from the origin diverges without resetting but is finite under resetting at a finite rate. 

\begin{figure}[t!]
    \centering
    \includegraphics[width=\linewidth]{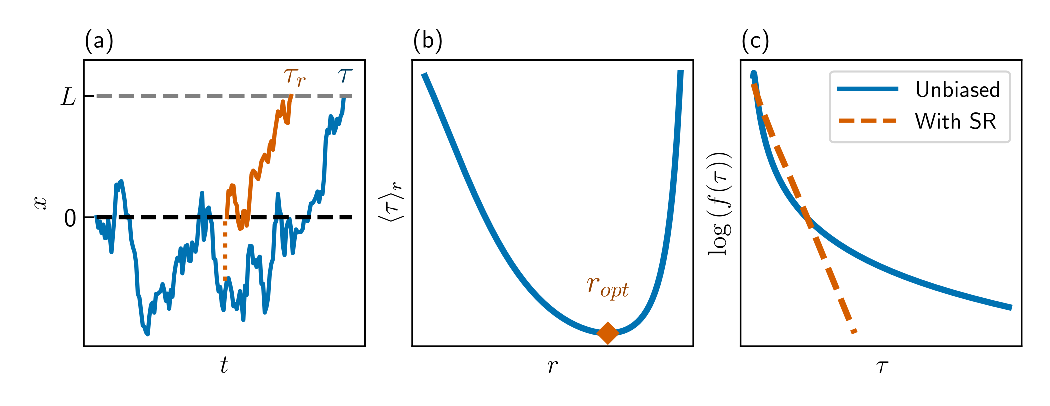}
    \caption{(a) An illustration of a one-dimensional diffusion from some initial position (black dashed line) to an absorbing boundary (gray dashed line). A continuous trajectory with FPT $\tau$ is plotted in blue. A trajectory with resetting follows the blue curve until it is restarted, as indicated by the orange dotted line. It then follows the orange plot and reaches the boundary at time $\tau_r$. (b) An illustration of a typical $\langle \tau \rangle_r$ as a function of $r$, with a minimum (orange rhombus) at resetting rate $r_{opt}$. (c) An illustration of FPT distributions with and without resetting, in orange and blue, respectively.}
    \label{fig:explainSR}
\end{figure}

Figure \ref{fig:explainSR}(a) illustrates the setup of Evans and Majumdar. The blue line shows a random trajectory without resetting along a coordinate $x$. The particle is initiated at the black dashed line, and first reaches the target gray dashed line located at $L$ at time $t=\tau$. We will refer to the time in which the system first reaches the target (with or without resetting) as the first-passage time (FPT). Alternatively, under resetting, the particle returns to the initial position at some random time, as indicated by the dotted orange line. In this specific example, the trajectory is composed of two independent segments separated by a resetting event, but it can generally be composed of any number of segments. The first segment follows the blue line until the restart and the second segment follows the orange solid line until reaching the target at time $\tau_r$. Note that $\tau_r$ is the cumulative duration of all segments in the trajectory under resetting. In this particular case, the total time it took the particle to reach the absorbing boundary is shorter with resetting, i.e., $\tau_r<\tau$. Evans and Madjumdar showed that for a diffusing Brownian particle introducing resetting at a small finite rate makes the mean FPT (MFPT) under restart finite. Therefore, 
$\langle \tau \rangle_r \ll \langle \tau \rangle$ for all finite $r$.

As we will discuss in the next section, resetting will not always accelerate a stochastic process. However, this powerful result, that introducing resetting lowers the MFPT, was found to be widespread, as demonstrated in various recent applications of resetting for accelerating stochastic processes: from queuing systems~\cite{bressloff_queueing_2020,bonomo_mitigating_2022} and randomized computer algorithms~\cite{gomes_boosting_1998,montanari_optimizing_2002}, to first-passage and search processes~\cite{kusmierz_optimal_2015,bhat_stochastic_2016,chechkin_random_2018,ray2019peclet,robin_random_2019,evans_run_2018,pal_search_2020,bodrova_resetting_2020,luo_anomalous_2022,yin2023restart}, and recently, by us, in enhanced sampling of MD simulations~\cite{Blumer2022,Blumer2024,non_exponential_kinetics,church2024acceleratingmoleculardynamicsinformed}. 
Apart from applications, resetting also emerged as a theoretical framework for a better understanding of biological phenomena such as enzymatic catalysis and inhibition~\cite{Michaelis_Menten_PNAS, Robin2018}, and backtrack recovery by RNA polymerases~\cite{Backtrack_RNA, Particle_evap}. In addition, its ability to generate non-trivial non-equilibrium steady states recently encouraged novel experimental studies in non-equilibrium statistical mechanics~\cite{altshuler2023environmental,Vatash2025,Tal-Friedman2020}. The ability to reduce the MFPT of search processes was also studied experimentally~\cite{Tal-Friedman2020, PRR_exp_reset, Faisant_2021_Exp}.

Figure \ref{fig:explainSR}(b) shows a typical example of the MFPT of processes accelerated by resetting as a function of the resetting rate $r$. If introducing resetting at a small finite rate is guaranteed to lead to acceleration, the MFPT first decreases with $r$, until reaching a minimum at some optimal resetting rate $r_{opt}$. It then must increase with $r$ since in the limit of $r \to \infty$ the system does not move at all, and we expect $\langle\tau\rangle_r$ to diverge. While Figure \ref{fig:explainSR}(b) shows the typical behavior, there are other cases, e.g., where the MFPT first increases with $r$, reaches a local maximum, and decreases at higher rates~\cite{Blumer2024,PhysRevE.92.060101,PhysRevResearch.1.032001}.

Resetting is particularly useful when the FPT distribution $f(\tau)$ has a slowly decaying tail. To better understand how resetting lowers the MFPT, observe the FPT distribution plotted in blue in Figure \ref{fig:explainSR}(c). While there is a high probability of sampling short FPTs, the distribution has a very broad tail (notice that it is plotted on a semi-logarithmic scale). Resetting eliminates the trajectories with extremely long FPTs and the FPT distribution decays much faster (orange), leading to a smaller MFPT.

\section{Theory of stochastic resetting}
\label{sec:theory}

In this section, we cover some of the theory of resetting we consider most relevant for enhanced sampling, and particularly the conditions underwhich a process is accelerated by resetting.

\subsection{Standard resetting}

The first question a practitioner might ask is whether resetting would accelerate a stochastic process. 
In the context of enhanced sampling this would mean reducing the MFPT for transitions between two metastable states separated by a free energy barrier. 
Fortunately, Reuveni et al. derived a universal sufficient condition to answer such questions, requiring little knowledge of the first-passage process~\cite{pal_inspection_2022,Evans_majumdar_JPhysA_review}. The first two moments of the FPT distribution determine whether resetting would be beneficial: if the standard deviation $\sigma$ is larger than the mean $\mu$, introducing a small finite resetting rate is guaranteed to lower the MFPT for any stochastic process. This condition is often written in terms of the coefficient of variation (COV),
\begin{equation}
    \text{COV} = \frac{\sigma}{\mu} > 1,
\end{equation}
which is known as the COV condition, or the COV test. As it only depends on the first two moments, a few samples of first-passage events without resetting provide an adequate estimation of the COV, and are enough to assess whether resetting should be used. 

Next, we would like to know what resetting rate to use. We summarize results for the two best-studied protocols of resetting. The first is Poisson resetting, where time durations $T$ between resets are taken from an exponential distribution with rate $r$, 
$f(T)=r\exp(-rT)$. The MFPT under Poisson resetting is related to the FPT distribution without resetting $f\left( \tau \right)$ 
through~\cite{reuveni_optimal_2016}
\begin{equation}
  \langle \tau \rangle_r = \frac{1-\tilde{f}(r)}{r\tilde{f}(r)}.
  \label{eq:MFPTpoisson}
\end{equation}
Here, $\tilde{f}(r)$ is the Laplace transform of $f\left( \tau \right)$,
\begin{equation}
    \tilde{f}(r) = \int_0^\infty e^{- r \tau} f\left( \tau \right) \, \mathrm{d}\tau = \langle e^{-r \tau } \rangle \approx
\frac{1}{N}\sum\limits_{i=1}^N e^{-r \tau_i},
\label{eq:Laplace}
\end{equation}
which, in the context of enhanced sampling, is estimated numerically from $i=1,2,\dots,N$ FPT samples $\tau_i$.

A second common resetting protocol is sharp resetting, where the time duration $T=1/r$ between restarts is constant.
The MFPT under sharp resetting is~\cite{Eliazar_sharp_2020}
\begin{equation}
  \langle \tau \rangle_T = \frac{1}{1-S(T)} \int \limits_0^T S(t) \mathrm{d} t,
  \label{eq:MFPTsharp}
\end{equation}
where $S(t)$ is the survival function of the process without resetting,
\begin{equation}
  S(t) = 1 - \int \limits_0^t f(\tau) \mathrm{d}\tau \approx 1 - \frac{\mathcal{N} \left\{\tau | \tau < t\right\}}{N},
\end{equation}
with $\mathcal{N} \left\{\tau | \tau < t\right\}$ being the number of samples for which $\tau_i < t$.
Thus, given some samples of first-passage events without resetting, one can estimate the MFPT under resetting for any resetting rate and assess the optimal rate, providing the highest acceleration. It is worth noting that sharp resetting is guaranteed to provide acceleration equal to or higher than any other resetting protocol when performed using the optimal resetting timer $T$~\cite{pal_first_2017}.

We acknowledge that sampling first-passage events without resetting for transitions between metastable states that are separated by high free energy barriers can be very challenging for standard MD simulations, due to the long timescales. When considering applying resetting as a standalone method, we assume some trajectories are already available to estimate the COV.

In addition, one can sample trajectories with an 
initial guess of resetting rate $r^*$. An approach that we found useful~\cite{Blumer2022} is to start with a rate that is on the same order of magnitude, but slighty larger, than a rough estimate of the reciprocal of the MFPT. It often provides some acceleration, and we then optimize the resetting rate in the following way: In the case of Poisson resetting, one first evaluates the Laplace transform of the FPT distribution $f_{r^*}(\tau)$ under resetting rate $r^*$ through Equation \ref{eq:Laplace}. Then, the MFPT under any resetting rate $r=r^*+\Delta r$ is given by a generalization of Equation \ref{eq:MFPTpoisson}~\cite{Blumer2022},
\begin{equation}
  \langle \tau \rangle_r = \frac{1-\tilde{f}_{r^*}(\Delta r)}{\Delta r\tilde{f}_{r^*}(\Delta r)}.
  \label{eq:MFPTpoissonDelta}
\end{equation}
In the case of sharp resetting, we note that the segments between restarts exactly sample the survival function without resetting up to the chosen time duration $T^*$. As Equation \ref{eq:MFPTsharp} only requires the survival up to the resetting time, one can use the sampled trajectories to assess the MFPT for any timer $T < T^*$.
Both approaches predict whether using higher rates than the initial guess will lead to higher acceleration without additional sampling.

Finally, we stress that all the equations above hold for a general stochastic process, and particularly simulations which employ enhanced sampling techniques. In that case, estimating the COV is easier. Given a few samples, even with suboptimal collective variables, one can estimate the COV and use Equations \ref{eq:MFPTpoisson} and \ref{eq:MFPTsharp} to find a beneficial resetting rate, with zero computational cost. As we will explain in detail in Section \ref{sec:SR_with_MetaD}, we believe a combined approach is often the most useful approach~\cite{Blumer2024}. Detailed examples of all of these use cases will be given in Setion~\ref{sec:appl}.

\subsection{Adaptive resetting}
\label{sec:ISR_theory}

In standard resetting, the resetting times are independent of the progress of the stochastic process. However, higher accelerations can be achieved by incorporating information on the state of the system into the resetting rate, e.g., by measuring the distance from the target state in CV space~\cite{church2024acceleratingmoleculardynamicsinformed,tal2025smart}. We call such state- and, possibly, time-dependent resetting protocols adaptive resetting. By employing an adaptive resetting rate $r\left(\boldsymbol{X}\right)$, one can avoid undesired restarts when the system is close to the target, or encourage frequent restarts when getting stuck in undesirable metastable states away from the target. Here and onward, $\boldsymbol{X}$ stands for the coordinates of the system, either Cartesian or in a low-dimensional CV space. 

The key difficulty in employing state-dependent resetting is that the general theory of SR we just reviewed does not hold for state-dependent resetting. It stems from the fact that, unlike in standard resetting, samples of the FPT distribution are insufficient, and the trajectories themselves, i.e., the progression of $\boldsymbol{X}$ in time, are required. 
We recently developed a general framework to estimate any observable of the process with adaptive resetting, given trajectories without resetting~\cite{keidar2024adaptive}.
When using simulations, the trajectories are readily available without any additional cost apart from computer memory for storing them. We denote the coordinates of trajectory $i$ of $n_i$ steps at timestep $j=1,\dots,n_i$ as $\boldsymbol{X}_i^j$.

The framework is valid for any state- and time-dependent resetting rate, but, for clarity, we focus on predicting the MFPT under an adaptive resetting protocol in which we allow Poisson resetting at rate $r$ only if the system is in some region $A$ of phase space. At any single timestep, the probability of resetting is
\begin{equation}
    p\left(\boldsymbol{X}\right) = r\Delta t \cdot h_A \left(\boldsymbol{X}\right),
    \label{eq:resetPro}
\end{equation}
where $\Delta t$ is the simulation time step and $ h_A \left(\boldsymbol{X}\right)$ is the indicator function,
\begin{equation}
h_A \left(\boldsymbol{X}\right) =\begin{cases} 1 & \text{if } \boldsymbol{X} \in A \\ 0 & \text{otherwise.} \end{cases}
\end{equation}
With this, we can evaluate the survival function of each sampled trajectory, that is, the probability it would have survived $k$ steps without restart if we had applied adaptive resetting,
\begin{equation}
    S_i (k) = \prod_{j=1}^{k-1} \left[ 1- p\left(\boldsymbol{X}_i^j\right) \right].
    \label{eq:survival}
\end{equation}

With $S_i (k)$ at hand for all $N$ trajectories, we can estimate three desired quantities of trajectories with resetting, as weighted averages of the available trajectories without resetting. These are the mean number of segments in a trajectory with resetting $\langle M \rangle_r$, the mean duration between resetting events $\langle T \rangle_r$, and the mean duration between the last resetting event and first-passage, $\langle T^f\rangle_r$~\cite{church2024acceleratingmoleculardynamicsinformed}:
\begin{equation}
    \begin{split} 
      &  \langle M\rangle_r=\frac{1}{\langle S \rangle}\approx \frac{1}{\frac{1}{N} \sum_{i=1}^N S_i(n_i)},\\
      &  \langle T \rangle_r=\frac{\langle(1-S)T\rangle}{\langle1-S\rangle}\approx\frac{\sum_{i=1}^N\sum_{j=1}^{n_i-1}S_i(j)p(\boldsymbol{X}_i^j)j\Delta t}{\sum_{i=1}^N \left( 1 - S_i(n_i) \right)},\\
      &  \langle T^f \rangle_r = \frac{ \langle \tau S \rangle}{\langle S \rangle} \approx \frac{ \sum_{i=1}^N \tau_iS_i(n_i)}{\sum_{i=1}^N S_i(n_i)}.
    \end{split}
\end{equation}
A trajectory with resetting is composed of $M$ segments: $M-1$ segments ending in restart, and a single segment ending in first-passage.
The MFPT with state-dependent resetting is thus simply
\begin{equation}
    \langle \tau \rangle_r = \left( \langle M \rangle_r -1 \right) \langle T \rangle_r +  \langle T^f \rangle_r.
    \label{eq:prediction}
\end{equation}
In practice, Equations \ref{eq:resetPro}-\ref{eq:prediction} can be implemented in a few Python code lines, rapidly predicting the MFPT under any adaptive resetting protocols given samples of $\boldsymbol{X}_i^j$. The explicit form of the adaptive resetting protocol enters only in Eq.~\ref{eq:resetPro} and should be changed accordingly.
As in standard Poisson resetting, we can use trajectories with adaptive resetting at rate $r^*$ to evaluate the MFPT at any rate $r = r^* + \Delta r$. To do so, we simply treat the coordinates $\boldsymbol{X}_i^j$ of the trajectories with resetting as described above for unbiased trajectories and substitute $r = \Delta r$  in Equation \ref{eq:resetPro}. 

\section{Applications of resetting in molecular simulations}
\label{sec:appl}

\subsection{Resetting as a standalone method}

We established that standard resetting accelerates first-passage processes for which the COV of the FPT distribution is larger than unity. In which processes should we expect it to happen? To gain intuition, we next describe several systems where such broad FPT distributions emerge, and resetting proved useful. 

\begin{figure}[t!]
    \centering
    \includegraphics[width=\linewidth]{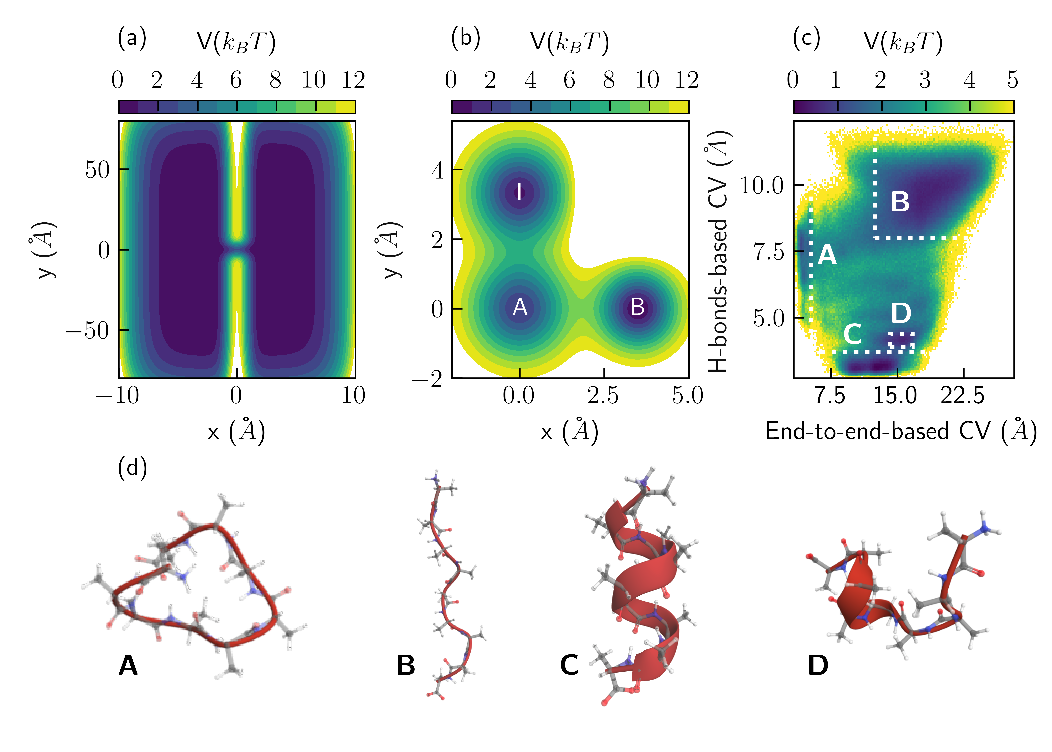}
    \caption{(a) The two-dimensional modified Faradjian-Elber potential. (b) The two-dimensional three-state potential. (c) Free-energy of a solvated alanine peptide along two CVs, based on the end-to-end distance or on intramolecular H-bonds. (d) Representative configurations of four metastable states of a solvated alanine peptide. The crimson cartoon represents the backbone of the peptide, while the white, gray, blue, and red spheres represent hydrogen, carbon, nitrogen, and oxygen atoms, respectively.}
    \label{fig:standAloneModels}
\end{figure}

Resetting is particularly efficient in flat landscapes~\cite{ray2019peclet,ray_diffusion_2020,ray_resetting_2021}. We expect it to accelerate processes where the free energy barriers are moderate and entropy plays a significant role in the dynamics. Then, the system explores nonreactive regions of phase space for long periods, resulting in trajectories with long FPTs. Occasionally resetting to a reactive region, where rapid transition to product is possible, can reduce the MFPT. For example, consider the modified Faradjian-Elber potential of Figure \ref{fig:standAloneModels}(a)~\cite{Blumer2024,faradjian_computing_2004,church2024acceleratingmoleculardynamicsinformed}. It is a two-dimensional model, consisting of two metastable states separated by an energy barrier of $12 \, k_B T$ at $x=0$ for most $y$-coordinate values, where $k_B$ is the Boltzmann constant and $T$ is the temperature. It has a narrow saddle only $3 \, k_B T$ high at $y=0$, near the minima of the basins. The basins are very broad, such that a diffusing particle starting from one of the minima may spend much time in nonreactive regions away from the saddle ($|y| \gg 0$). The COV for transitions from one minimum to the other basin is $\sim 1.23$. Poisson resetting at the optimal resetting rate gives a speedup of $\sim 15$~\cite{Blumer2024}, which is defined as the ratio of the MFPT without and with SR.

Often, barriers are much higher than $k_B T$. In a classical two-state scenario with a high barrier as in the Kramers' problem~\cite{KRAMERS1940284}, the FPT distribution is exponential: the COV is exactly 1 and resetting cannot accelerate transitions. However, this simplified picture often only holds at long timescales. When the initial state is composed of several sub-states, the rate of transitions can vary before reaching local equilibrium within the initial state. This is demonstrated in the three-state model of Figure \ref{fig:standAloneModels}(b)~\cite{non_exponential_kinetics,khan2020fluxional}. We initiate the particle in state A and are interested in transitions over a $\sim 10 \, k_B T$ barrier to the global minimum of state B. The particle can also overcome a $\sim 6 \, k_B T$ barrier to reach a nonreactive state I. The system reaches thermal equilibrium between states A and I after $\sim 1.5 \, \text{ns}$, after which the FPT distribution decays exponentially, with a MFPT of $\sim 130 \, \text{ns}$. But, before reaching equilibrium, the rate of events is faster, and sharp resetting with a timer of $1.15 \, \text{ns}$ gives a speedup close to $6$ over simulations without resetting~\cite{non_exponential_kinetics}.

We conclude that SR is efficient as a standalone method when there are multiple metastable states, or when the energy barriers are relatively low. Proteins are characterized by rugged free energy surfaces that often fit these criteria~\cite{MOULICK2019807,nevo2005direct,wolynes1996fast}. As an example, Figure \ref{fig:standAloneModels}(c) presents the free energy surface of a nine-residues alanine peptide in solution, along two CVs: the end-to-end distance and the mean distance of three H-bonds-forming sites~\cite{ayaz2021non}. Notice that the scale of the color bar is different from that in panels (a) and (b). We recognize 4 metastable states, marked with white dashed lines. Representative configurations of the states are given in Figure \ref{fig:standAloneModels}(d). Transitions from state A to any of the others (B, C, or D) have a MFPT of $\sim 100 \, \text{ns}$ and a COV of $\sim 1.46$. Sharp resetting accelerates this process by an order of magnitude~\cite{non_exponential_kinetics}.

\subsection{Combining resetting with MetaD}
\label{sec:SR_with_MetaD}
Resetting leads to substantial acceleration for entropic barriers, but is not always sufficient as a standalone method. It is less efficient for high enthalpic barriers in narrow basins, such that the FPT distribution is exponential. Most enhanced sampling methods are particularly designed to tackle such cases but, while being very successful, they have other drawbacks. Combining them with Resetting can compensate for these drawbacks, and lead to higher acceleration than either approach separately, as we explain below for the example of MetaD.

MetaD accelerates slow processes by introducing an external bias potential along chosen CVs. The bias is deposited on-the-fly at visited regions of the low-dimensional CV space, encouraging the exploration of poorly visited areas. In the popular well-tempered version~\cite{barducci_well-tempered_2008}, the bias is guaranteed to converge, such that at long times the effective free energy surface in the simulation is $\frac{1}{\gamma} F(s)$, with $F(s)$ being the true free energy along CV $s$ and $\gamma > 1$, called the bias factor, is a control parameter. However, the time for convergence is unknown a priori and can be long. Furthermore, the efficiency highly depends on the choice of CV: if $F(s)$ poorly reflects the slowest modes of the system, acceleration of rare transitions is limited.

\begin{figure}[t!]
    \centering
    \includegraphics[width=\linewidth]{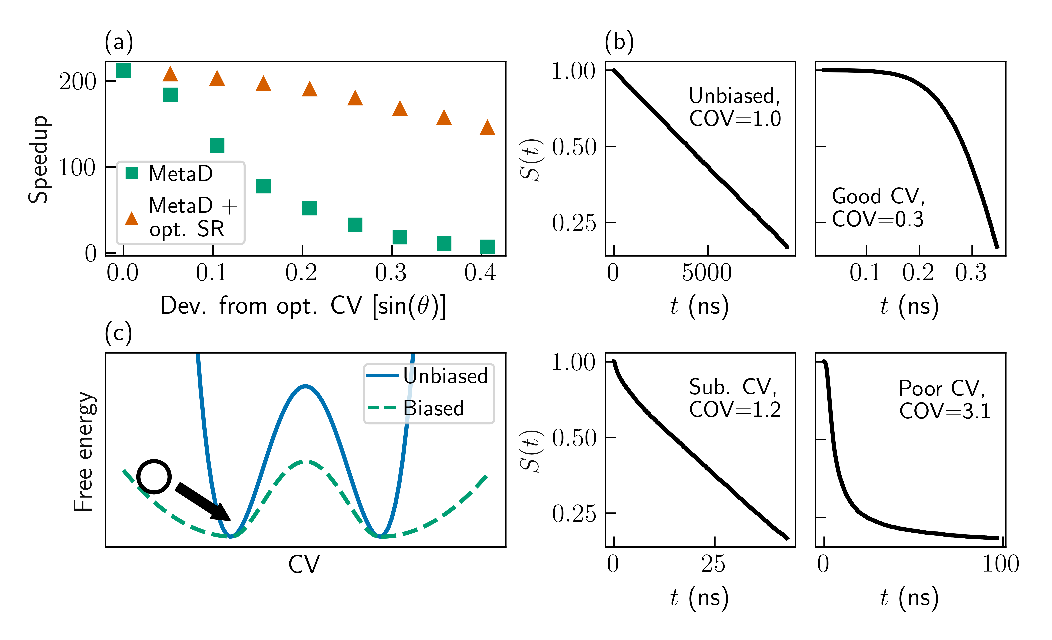}
    \caption{(a) Speedup in MetaD simulations of the two-dimensional modified Faradjian-Elber potential as a function of the CV quality. The quality is measured by the rotation angle relative to the optimal CV. Results are given either with or without resetting, in orange triangles and green squares, respectively. (b) Survival functions $S(t)$ for transitions between an ``unfolded" and ``folded" conformers of alanine tetrapeptide, in unbiased simulations (upper left) and MetaD simulations with a good CV ($\phi_3$ dihedral angle, upper right), a suboptimal CV ($\psi_3$, lower left), or a poor CV ($\phi_2$, lower right). (c) An illustration of the effect of MetaD bias on the effective free energy. Representative unbiased and biased FESs are given in blue and green, respectively. The black hollow circle represents a system on the biased surface, far from the energy barrier.}
    \label{fig:metad}
\end{figure}

Figure \ref{fig:metad}(a) demonstrates the dependency of the acceleration on CV choice through the example of the modified Faradjian-Elber potential (Figure \ref{fig:standAloneModels}(a)). There, the optimal CV is the x-axis. In Ref.~\cite{Blumer2024}, we artificially degraded the quality of the CV by rotating it at an angle $\theta$ with respect to the x-axis and measured first-passage from the right minimum to the left basin ($x<-1\, \AA$). The speedup as a function of $\theta$ is plotted with green squares. It is $> 200$ with the optimal CV but decreases with increasing $\theta$, and is close to 1 with poor CVs. For each CV, we added Poisson resetting at the optimal resetting rate on top of MetaD and plotted the speedups with orange triangles. The speedup decreases with $\theta$ at a much slower rate, and even poor CVs provide speedups close to that of the optimal CV. This result shows that SR can be an alternative or complementary method to CV optimization.

Ref.~\cite{Blumer2024} gives two molecular examples that show the same qualitative behavior. First, the transition between two metastable states of alanine tetrapeptide was accelerated using three different CVs, based on internal dihedral angles. For definitions of the angles, see Ref.~\cite{Blumer2024}. The $\phi_3$ dihedral angle, which is the optimal CV for the transition, provides a speedup of more than four orders of magnitude, and resetting does not give additional acceleration. The $\psi_3$ angle, a suboptimal CV, provides a speedup by two orders of magnitude, which is doubled when adding SR at the optimal resetting rate. Lastly, $\phi_2$ is an extremely poor CV, providing a speedup of less than an order of magnitude without SR. But, by resetting, the speedup becomes close to that achieved by $\psi_3$. As another example, the folding of the chignolin mini protein (Figure \ref{fig:chignolin}) was enhanced using two CVs. A CV based on harmonic linear discriminant analysis~\cite{mendels2018folding} provided speedups of two orders of magnitude, and resetting further increased the speedup by a factor of $\sim 1.5$. A CV based on the C-alpha root-mean-square deviation (RMSD) from a folded configuration provided speedups of $\sim 10$ without resetting, but $> 100$ with SR. 

To better understand the cause of this trend, we plot the survival function for alanine tetrapeptide in Figure \ref{fig:metad}(b). The unbiased FPT distribution (top left) is indeed exponential (Note that the survival function of an exponential distribution is linear on a semi-log scale). The COV=1, and resetting cannot be used as a standalone method. However, the bias of MetaD disturbs the dynamics and changes the FPT distribution dramatically. Biasing along $\phi_3$, the distribution decays much faster than exponential, and the COV is $<1$ (top right). With $\psi_3$, the distribution has an exponential tail but decays faster at short times, and the COV=1.2 (bottom left), which leads to acceleration by resetting. With the poorest CV, $\phi_2$, the distribution is much broader (notice the different range of the x-axis) and we find COV=3.1, explaining the great acceleration by resetting in this case.

We expect resetting to have a similar influence on additional CV-based enhanced sampling methods apart from MetaD. The reason is that CV-based methods accelerate the exploration of phase space by flattening the free energy surface. While this lowers energy barriers, resulting in accessible transitions between metastable states, it can also encourage sampling of high-energy regions of phase space away from the barrier, which are often not interesting or even nonphysical. Simulations with suboptimal CVs are particularly susceptible to this. Resetting can help avoid this by returning the system to local minima. As resetting is known to be particularly useful for flat landscapes, it naturally compensates for this drawback.

We note that when resetting MetaD simulations, all the bias is zeroed. In addition, we emphasize that, when combining SR with enahnced sampling methods, it is much easier to know whether resetting is favorable and what rates to use than for the standalone case. Since transitions are fast in comparison to unbiased simulations, it is easier to sample them without resetting and estimating the COV. Given several trajectories without resetting, a practitioner may evaluate the MFPT under any resetting rate through Equations \ref{eq:MFPTpoisson} or \ref{eq:MFPTsharp}, with no computational cost. Resetting can then be used to expedite additional sampling.

\subsection{Informed resetting}

\begin{figure}[t!]
    \centering
    \includegraphics[width=\linewidth]{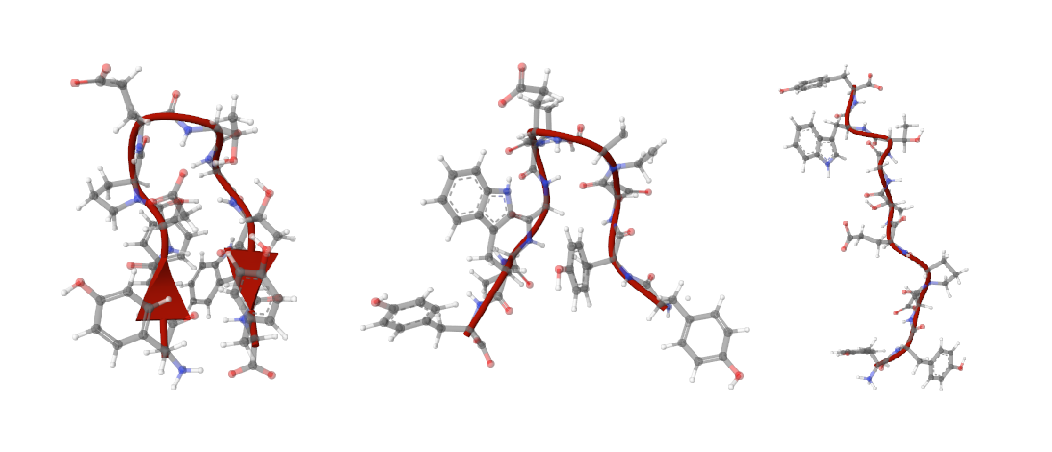}
    \caption{Representative configurations of three metastable states of the chignolin mini protein -- folded, misfolded, and unfolded (from left to right, respectively). The crimson cartoon represents the backbone of the protein, while the white, gray, blue, and red spheres represent hydrogen, carbon, nitrogen, and oxygen atoms, respectively.}
    \label{fig:chignolin}
\end{figure}

In standard SR, the timing of resetting events is completely decoupled from the state of the system. It is one of the advantages of resetting as an enhanced sampling method: limited prior knowledge is needed, and no CVs are required. However, it also means that restarts sometimes occur very close to the target. Greater accelerations can clearly be achieved by avoiding such undesirable restarts. 
In informed stochastic resetting (ISR), we exploit information on the progress of the process by incorporating an adaptive, state-dependent resetting protocol, resulting in higher speedups.

Ref.~\cite{church2024acceleratingmoleculardynamicsinformed} first demonstrated the potential of informed resetting for accelerating molecular simulations. There, a simple form of state-dependent resetting was considered: when it is time for a resetting event, the system is restarted only if the value of a chosen CV is greater than some predefined threshold. We first studied the modified Faradjian-Elber potential of Figure \ref{fig:standAloneModels}(a).
Using ISR as a standalone method, and choosing the x-axis as CV, provided a speedup of $>50$ over simulations without resetting. Using the same CV and threshold, but combining ISR with MetaD, gave a speedup of $\sim 700$. Even when using a very poor CV (corresponding to the rightmost points in Figure \ref{fig:metad}(a)), the speedup remained $>200$.

We also employed informed resetting to accelerate conformational changes in chignolin in explicit water. We focused on transitions from a misfolded state to the native, folded state. Most trajectories first unfold before reaching the native state, but some rapidly reach it directly from the misfolded state 
(representative configurations of the folded, misfolded, and unfolded states are given in Figure \ref{fig:chignolin}). By resetting only when unfolding, the MFPT is decreased by $15 \,\%$ even without MetaD. Introducing bias along the RMSD-based CV, MetaD gives speedups of $\sim 60$ when used as a standalone method. Adding standard SR increases the speedup up to $\sim 300$, but ISR increases it up to $\sim 950$.

For this molecular system, we employed a simple informed resetting protocol: no resetting for RMSD values smaller than some threshold, and Poisson resetting at some finite rate otherwise. In practice, however, the resetting rate can be any function of the spatial and momenta coordinates or some low-dimensional CV space. In Ref.~\cite{keidar2024adaptive}, we showed that a neural network-based resetting protocol relying on three CVs (the RMSD, the radius of gyration, and the end-to-end distance) could provide greater speedups than the one-dimensional threshold-based protocol employed in Ref.~\cite{church2024acceleratingmoleculardynamicsinformed}. Moreover, we found neural network-based resetting protocols that accelerate transitions from the unfolded state to the native state -- a process for which one-dimensional thresholds along the RMSD did not improve over standard resetting. This proves that incorporating additional information into the resetting scheme can provide further acceleration. But, as is common in CV-based methods, the performance of ISR depends on the correct utilization of the available information. When prior knowledge is lacking, one may better begin by sampling trajectories with standard SR. Insights from those trajectories can later be used to design informed resetting protocols for additional sampling with improved efficiency.

\section{Kinetics inference}
\label{sec:kinetics}

We now discuss another major goal of enhanced sampling: inferring the unbiased kinetics from accelerated simulations. We first consider the inference of the MFPT. We overview two approaches designed for SR as a standalone method, and a suitable approach when incorporating SR in MetaD simulations. The latter is also appropriate for improving the efficiency of kinetics inference from MetaD simulations without resetting~\cite{blumer2024shorttime}. Finally, we discuss the inference of kinetic properties other than the MFPT, focusing on direct transit times.

\subsection{Inference of the MFPT for SR as a standalone method}

In the first approach, we obtain the MFPT under resetting at different rates $r$ and fit a function to $\langle \tau \rangle_r$. The estimation of the unbiased MFPT is simply the value of the fitted function at $r=0$. Different techniques can be employed for the fit. In Ref.~\cite{Blumer2022}, we obtained $\langle \tau \rangle_r$ at nine equally-spaced grid points and fitted them with a fourth-order Taylor series, using finite-difference methods to evaluate the derivatives. Crucially, performing many simulations to obtain the MFPT at various rates would have been prohibitive. Luckily, instead, we only need to perform simulations at a single resetting rate $r^*$ and obtain the MFPT at higher rates as explained in Section \ref{sec:theory}.

\begin{figure}[t!]
    \centering
    \includegraphics[width=\linewidth]{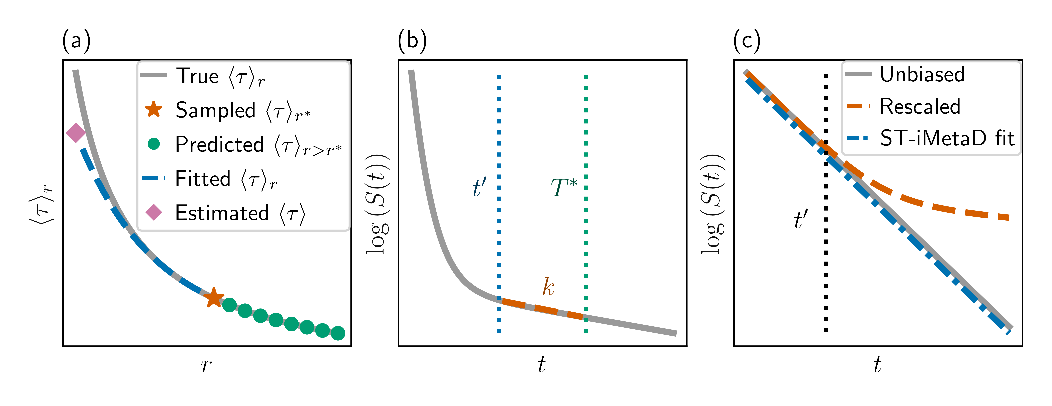}
    \caption{(a) An illustration of MFPT inference procedure through fitting $\langle \tau \rangle_r$. The MFPT at simulated resetting rate $r^*$, the predicted MFPT at $r>r^*$, the fit of $\langle \tau \rangle_r$, and the estimated unbiased MFPT are presented with an orange star, green circles, blue dashed line, and pink rhombus, respectively. The true $\langle \tau \rangle_r$ is plotted with a gray line. (b) An illustration of a survival function (gray line) with an exponential decay with rate $k$ at $t>t'$, with $t'<T^*$. Times $t'$ and $T^*$ are marked with blue and green dashed lines, respectively. A linear fit to $\log\left(S(t)\right)$ at $t'<t<T^*$ is plotted with an orange dashed line. (c) An illustration of a rescaled survival function in MetaD simulations (orange) that deviates from the unbiased survival (gray). A linear fit (blue) of $\log\left(S(t)\right)$ at times $t < t'$ (marked with a black dotted line) follows the unbiased survival function.}
    \label{fig:kinetics}
\end{figure}

This inference procedure is illustrated in Figure \ref{fig:kinetics}(a): simulations with resetting rate $r^*$ directly provide $\langle \tau \rangle_{r^*}$ (orange star), and the FPT distribution under resetting rate $r^*$, $f_{r^*}(\tau)$. With $f_{r^*}(\tau)$ at hand, the MFPT under $r>r^*$ is predicted through Equations \ref{eq:MFPTsharp} or \ref{eq:MFPTpoissonDelta} (green circles). A function is fitted (blue dashed line) and extrapolated to $r \to 0$ to obtain the unbiased MFPT, $\langle\tau\rangle_0$ (pink rhombus). We note that this approach is also applicable to simulations with ISR as a standalone method~\cite{church2024acceleratingmoleculardynamicsinformed}.

The second approach is suited for sharp resetting at constant time intervals $T^*$. We first use the total expectation theorem to express the unbiased MFPT as 
\begin{equation}
\langle \tau \rangle = \left(1-S(T^*)\right) \langle \tau | \tau \le T^* \rangle + S(T^*) \langle \tau | \tau > T^* \rangle,
\label{eq:trueMFPT}
\end{equation}
where $\langle \tau | \tau \le T^* \rangle$ is the conditional MFPT of trajectories with FPT $<T^*$ and $\langle \tau | \tau > T^* \rangle$ is the conditional MFPT of trajectories with FPT $>T^*$. We note that both $S(T^*)$ and $\langle \tau | \tau \le T^* \rangle$ are directly sampled in simulations with sharp resetting.

To evaluate $\langle \tau | \tau > T^* \rangle$, we assume the survival function has some distinctive tail at times larger than some $t'<T^*$, as illustrated in Figure \ref{fig:kinetics}(b).
Since simulations with sharp resetting sample the unbiased survival at times $< T^*$, the tail is sampled without bias at times $t'<t<T^*$. We evaluate $\langle \tau | \tau > T^* \rangle$ from the behavior of the tail: For instance, if we assume an exponential tail with rate $k$, we can extract $k$ from the survival function and estimate 
\begin{equation}
\langle \tau | \tau > T^* \rangle = T^* + k^{-1}.
\label{eq:trueMFPTexp}
\end{equation}
This can be easily done with a linear fit to the logarithm of the survival function, as illustrated in Figure \ref{fig:kinetics}(b).
As another example, if we assume a power-law decay, i.e, $S(t>t') \propto t^{-\alpha}$ with some $\alpha>1$, we can estimate 
\begin{equation}
\langle \tau | \tau > T^* \rangle = \frac{\alpha T^*}{\alpha-1}.
\label{eq:paretoPrediction}
\end{equation}
We can obtain $\alpha$ by a linear fit to $S(t)$ on a log-log scale.
Practically, we take $t'$ and the fitting parameters that fit the data best, and obtain the unbiased MFPT by substituting $\langle \tau | \tau > T^* \rangle$ in Equation \ref{eq:trueMFPT}.

\subsection{Short-time infrequent MetaD}

Sharp resetting is more suitable for kinetics inference from MetaD simulations with resetting. As in the case of sharp resetting of standard MD trajectories, we treat each segment between restarts as an independent simulation. We then analyze them using short-time infrequent MetaD~\cite{blumer2024shorttime} (ST-iMetaD), as we next explain. For clarity, we first discuss how ST-iMetaD is applied to MetaD simulations without resetting.

When inferring kinetics from MetaD simulations, several assumptions are usually made. Mainly, we assume that the underlying FPT distribution is exponential and that the external bias does not affect the transition state~\cite{tiwary_metadynamics_2013,palacio-rodriguez_transition_2022,ray_kinetics_2023,blumer2024shorttime,mazzaferro2024good}. When the assumptions are valid, the ratio of the unbiased MFPT to the effective MFPT with MetaD, $\langle \tau \rangle_M$, is~\cite{tiwary_metadynamics_2013,valsson_enhancing_2016,ray_kinetics_2023}
\begin{equation}
\frac{\langle \tau \rangle}{\langle \tau \rangle_M} \approx
\frac{Z}{Z_M} = \langle e^{\beta V}\rangle,
\end{equation}
where $Z$ and $Z_M$ are the partition functions confined to the initial state, with and without the external bias $V$, respectively, and $\beta$ is the inverse temperature. Practically, the FPT $\tau_i$ of each MetaD trajectory is rescaled to give $\tilde{\tau}_i = \alpha_i \tau_i$, with~\cite{valsson_enhancing_2016,ray_kinetics_2023} 
\begin{equation}
\alpha_i = \frac{1}{n_i}\sum \limits_{j=1}^{n_i} e^{\beta V_i(j)}.
\end{equation}
Here, $V_i(j)$ is the bias experienced in trajectory $i$ at timestep $j$. The total nuber of steps in trajectory $i$ is $n_i$. If the assumptions hold, the rescaled FPT distribution $f\left(\tilde{\tau}\right)$ should be exponential. As a result, in standard infrequent MetaD (iMetaD), an exponential fit is performed, and its mean is taken as the estimation for the unbiased MFPT~\cite{salvalaglio_assessing_2014,ray_kinetics_2023}.

However, in many cases, the assumptions break and $f\left(\tilde{\tau}\right)$ is far from exponential. The main causes for error are bias over-deposition and hysteresis, which may arise due to suboptimal choice of CV and too frequent bias deposition~\cite{salvalaglio_assessing_2014,blumer2024shorttime,palacio-rodriguez_transition_2022,ray_kinetics_2023,mazzaferro2024good}. 
Interestingly, we obsreved that $f\left(\tilde{\tau}\right)$ often follows the true $f\left(\tau\right)$ at short times, even when it dramatically deviates from it at long times~\cite{Blumer2024,blumer2024shorttime}. The reason is that trajectories with short $\tilde{\tau}$ experience relatively little bias, and are less vulnerable to the errors of bias over-deposition. Figure \ref{fig:metad}(c) illustrates it through the logarithm of the survival function, which is linear for the exponential distribution (gray line). The rescaled survival function (orange dashed line) shows the correct linear decay at rate $k$ up to some time $t'$, where it deviates.

In ST-iMetaD~\cite{blumer2024shorttime}, we limit the analysis to short timescales. We find a linear fit to $\log\left(S\left(\tilde{\tau}\right)\right)$ only at times $<t'$ to obtain $k$, and estimate the unbiased MFPT as $-k^{-1}$. This fit is illustrated with a blue dash-dotted line in Figure \ref{fig:kinetics}(c), showing good agreement with the unbiased survival. Practically, we take $t'$ and $k$ best matching the data. ST-iMetaD showed improved accuracy in comparison to standard iMetaD for predicting the unfolding rate of chignolin~\cite{blumer2024shorttime} and several protein-ligand residence times~\cite{LeeProteinLigandResidenceTimes} with enough samples.
Because ST-iMetaD does not require the entire rescaled FPT distribution, it is also suitable for simulations with resetting.  In that case, resetting does not bias the survival function at times shorter than the resetting rate and we take the short time fit of the survival function in a similar manner. Moreover, resetting minimizes the amount of long simulations with bias over-deposition, so they do not contribute to the ST-iMetaD fit.

\subsection{Direct transit times}

\begin{figure}[t!]
    \centering
    \includegraphics[width=\linewidth]{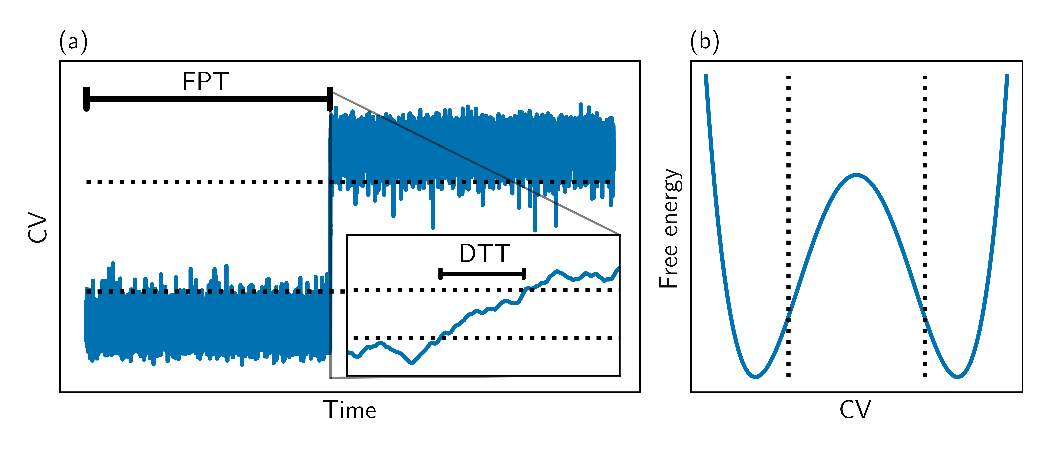}
    \caption{(a) A trajectory in a two-state system. The inlet zooms in on the DTT. (b) A schematic two-state system. The black dotted lines in both panels mark the thresholds that define the DTT.}
    \label{fig:DTT}
\end{figure}

When considering inference of the MFPT from biased simulations, there is always a tradoff between speedup and accuracy~\cite{Blumer2022,blumer2024shorttime,salvalaglio_assessing_2014,palacio-rodriguez_transition_2022,mazzaferro2024good}. Significant accelerations often carry errors of orders of magnitude in the MFPT estimation~\cite{Blumer2024}. However, there are other important kinetic properties that are less sensitive to bias and can be reliably assessed even at high accelerations. One such property is the direct transit time (DTT), which is the time duration between the last crossing of some predefined high-dimensional surface, to the first crossing of a second predefined surface~\cite{Elber_Makrov_Orland_Book}. It is often defined using two thresholds of some CV, centered around the transition state~\cite{Satija2020}, estimating the barrier-crossing time. DTTs were measured experimentally for biological macromolecules~\cite{SungChung2012, Neupane2016, Sturzenegger2018} and were studied both theoretically~\cite{Satija2020, Laleman2017} and computationally~\cite{Mori2016, Satija2017}. To better illustrate what is a DTT, Figure \ref{fig:DTT}(a) shows a typical trajectory in a two-state system along some CV. The system remains in the lower state for a long period of time, before showing a quick transition to the upper state. The DTT is the duration of just the transition between the two states (see inlet), which is significantly shorter than the FPT. It is defined by the black dotted lines, which are plotted against a schematic two-state free energy surface in Figure \ref{fig:DTT}(b).

Because DTTs are intrinsically shorter than FPTs, if the MetaD bias over the transition state is minimal, DTTs are much less vulnerable to inference errors that originate in bias over-deposition. They are also less sensitive to resetting at random times, since the time spent near the barrier is much shorter. 
In Ref.~\cite{church2024acceleratingmoleculardynamicsinformed}, we estimated the mean DTT for the folding of chignolin from accelerated simulations using either MetaD, MetaD + SR, and MetaD + ISR. 
All three methods exhibit similar errors, less than an order of magnitude, with speedups of up to 60, 300, and 950 for MetaD, MetaD + SR, and MetaD + ISR, respectively. In this
case, MetaD + ISR is the most advantageous, providing the highest acceleration with similar accuracy as the other methods.

\section{Looking forward}
\label{sec:lookout}

Standard and informed resetting showed promising results for enhanced sampling of molecular simulations, either as a standalone method or in combination with MetaD. In this section, we would like to discuss and suggest future developments that would broaden its capabilities.

First, resetting can be combined with other enhanced sampling methods besides MetaD.
If the COV under the influence of the enhanced sampling method is greater than unity, there is a finite resetting rate that will lower the MFPT. As we explained in Section \ref{sec:SR_with_MetaD}, we anticipate enhanced sampling methods that rely on the deposition of external bias, such as Gaussian accelerated MD~\cite{doi:10.1021/acs.jctc.5b00436}, Adiabatic free energy dynamics~\cite{rosso_adiabatic_2002,rosso_use_2002}, or OPES~\cite{invernizzi_rethinking_2020,invernizzi_opes_2021}, to benefit from resetting similarly to MetaD. It would be interesting to see how other approaches, such as replica exchange or transition path sampling, would benefit from future combinations.

Apart from acceleration, resetting provides inference of kinetic properties, such as the MFPT and the mean DTT. However, new methods need to be developed for the inference of thermodynamic properties, such as free energy differences or free energy surfaces. Inference methods for thermodynamic properties would be a welcomed addition to the toolset of the method, making it more versatile and attractive to use.
As a starting point, Poisson resetting is proven to lead to a non-equilibrium steady state, with a known relation to the propagator without resetting~\cite{Evans_majumdar_JPhysA_review}. 

Another promising direction for future development is informed resetting. In Ref.~\cite{church2024acceleratingmoleculardynamicsinformed}, we considered simple step-functions as resetting protocols. This approach did not exploit the whole potential of incorporating information, as the resetting rate can be \textit{any} function of \textit{all} degrees of freedom of the system. Together with the framework (see Equations \ref{eq:resetPro}-\ref{eq:prediction}) to estimate the MFPT under any adaptive resetting protocol given some sampled trajectories without resetting (Section \ref{sec:ISR_theory}), we envision automatic optimization of the resetting protocols, which can be updated on-the-fly with additional sampling. Very recently, we represented the resetting protocol by neural network and optimized it to minimize the MFPT using Equation \ref{eq:prediction} as a loss function~\cite{keidar2024adaptive}. A few other examples of learning how to reset have emerged recently in a different context~\cite{munozgil2025learningresettargetsearch}, and we anticipate this to be an exciting forefront of the field.

\section{Conclusions}

To conclude, resetting is an easy-to-use method that can be employed either as a standalone tool or in combination with any other enhanced sampling technique. It accelerates the sampling of slow first-passage events while providing accurate estimations of the first-passage kinetics. Standard resetting requires minimal prior knowledge of the studied system. However, when information about the progress of the process is available, state-dependent resetting protocols can provide greater acceleration.

The theory of resetting provides a powerful tool to analyze trajectories without resetting 
and decide: will resetting accelerate my sampling and by how much? What is the optimal rate, and how can I obtain additional acceleration by using smart adaptive resetting strategies?
For trajectories already accelerated by some algorithm of enhanced sampling, it is even easier to know whether resetting can be helpful. It requires zero computational effort, only the first moments of the FPT distribution in the case of standard resetting and some samples of trajectories for informed resetting.
Finally, resetting is trivially parallelizable. When considering parallel simulations, resetting has a great advantage: for instance, if we initiate simulations without resetting and stop them when observing first-passage, even a single trajectory with an extremely long FPT delays us. We can better exploit our resources by employing resetting, decreasing the real-time sampling duration by orders of magnitude.

\section*{Funding information}
O. B. acknowledges support from the Clore Scholars Programme of the Clore Israel Foundation. B. H. acknowledges support from the Israel Science Foundation (grants No. 1037/22 and 1312/22) and the Pazy Foundation of the IAEC-UPBC (grant No. 415-2023).

\section*{Conflicts of interest}
The authors declare no conflicts of interest.

\bibliography{main}

\end{document}